\documentclass[12pt]{article}
\usepackage{amsfonts,amssymb,mathrsfs}
\usepackage{hyperref}
\begin{document}
\begin{center}
{\Large\bf $AdS_4\times\mathbb{CP}^3$ superstring in the light-cone gauge}\\[0.5cm]
{\large D.V.~Uvarov\footnote{E-mail: d\_uvarov@\,hotmail.com, uvarov@\,kipt.kharkov.ua}}\\[0.2cm]
{\it NSC Kharkov Institute of Physics and Technology,}\\ {\it 61108 Kharkov, Ukraine}\\[0.5cm]
\end{center}

\begin{abstract}
The Type IIA superstring action on the $AdS_4\times\mathbb{CP}^3$
background, obtainable by the double dimensional reduction of the
$AdS_4\times S^7$ supermembrane, is considered in the
$\kappa-$symmetry light-cone gauge, in which the light-like
directions are chosen on the $D=3$ Minkowski boundary of  $AdS_4$.
Such choice of the gauge condition relies on representing the
$AdS_4\times S^7$ background isometry superalgebra $osp(4|8)$ (and
correspondingly the $osp(4|6)$ isometry superalgebra of the
$AdS_4\times\mathbb{CP}^3$ background) as $D=3$ extended
superconformal algebra. The gauge-fixed action includes
contributions up to the 4th power in the fermions.
\end{abstract}
\section{Introduction}

Proposed by Aharony, Bergman, Jafferis and Maldacena (ABJM)
\cite{ABJM} the duality relation between the $D=3$ $\mathcal N=6$
superconformal Chern-Simons-matter theory with the $U(N)\times
U(N)$ gauge symmetry and level $k$ and M-theory on the $AdS_4\times(
S^7/\mathbf Z_k)$ background provided the novel instance of the
gauge fields/strings correspondence. Whenever $N,k\to\infty$ with
the t'Hooft coupling $\lambda=N/k$ fixed the dual theory reduces
to the IIA superstring on the $AdS_4\times\mathbb{CP}^3$ background.
So one of the pivotal problems in exploring the ABJM duality is to
construct and quantize the string theory on
$AdS_4\times\mathbb{CP}^3$. The problem of constructing the
superstring action on $AdS_4\times\mathbb{CP}^3$ including the
fermions has been addressed in \cite{AF}-\cite{GSWnew}.  In Ref.
\cite{AF} there was applied the supercoset approach previously
proposed to obtain the Green-Schwarz (GS) superstring action on
the $AdS_5\times S^5$ superbackground \cite{MT}, \cite{Kallosh}. The
basic observation is that the symmetry group $SO(2,3)\times SU(4)$
of the bosonic background can be viewed as the bosonic subgroup of the
$OSp(4|6)$ supergroup that also includes 24 fermionic generators. The
action constructed in \cite{AF} using the $OSp(4|6)/(SO(1,3)\times
U(3))$ supercoset element includes 10 bosonic and 24 fermionic\footnote{Equal in number to the Type IIA supersymmetries
preserved by the $AdS_4\times\mathbb{CP}^3$ background.} degrees
of freedom, is invariant under the $8-$parameter $\kappa-$symmetry
transformations and is classically integrable. Such supercoset
action corresponds to the complete superstring action on
$AdS_4\times\mathbb{CP}^3$ with 32 fermions \cite{GSWnew}, in
which the $\kappa-$symmetry gauge freedom has been partially
fixed. Since $S^7$ admits the Hopf fibration representation as
$\mathbb{CP}^3\times U(1)$ \cite{Pope}, \cite{Volkov} the complete superstring action on
$AdS_4\times\mathbb{CP}^3$ can be constructed by performing the
double dimensional reduction \cite{DHIS} of the membrane action on the $D=11$
maximally supersymmetric background $AdS_4\times S^7$
\cite{Plefka98} that can also be viewed as the supercoset manifold
$OSp(4|8)/(SO(1,3)\times SO(7))$. This complete action on
$AdS_4\times\mathbb{CP}^3$ describes all possible superstring
motions and allows a wider choice of the $\kappa$-symmetry gauges
compared to the supercoset action. However, whether it is
integrable remains unclear.

Among possible $\kappa-$symmetry gauge conditions a special role
is played by the light-cone type gauges. In flat superspace it is
the light-cone gauge in which the GS equations of motion linearize and the model can be straightforwardly
quantized. Light-cone type gauge conditions also appear to be useful in exploring the string theory on $AdS_5\times S^5$ (see \cite{found} for review and
references). The
superstring model on $AdS_4\times\mathbb{CP}^3$, as well as its Penrose limit have
been studied in the light-cone type gauges in \cite{Nishioka}-\cite{Dukalski}. Here we consider the
superstring action on the $AdS_4\times\mathbb{CP}^3$ background
beyond the $OSp(4|6)/(SO(1,3)\times U(3))$ supercoset approach in the fermionic light-cone gauge corresponding to the choice of the light-like directions
within the $D=3$ Minkowski boundary of $AdS_4$ \footnote{Analogous $\kappa-$symmetry gauge conditions were proposed in \cite{MTlc} for the
$AdS_5\times S^5$ superstring.}. Similarly to our previous paper
\cite{0811} we elaborate on the presentation of the $AdS_4\times S^7$ background isometry superalgebra $osp(4|8)$ as the $D=3$ extended superconformal algebra
and the transformation of the Cartan forms to the
superconformal basis\footnote{The splitting of the fermionic generators/coordinates into the supersymmetry/superconformal ones w.r.t. the symmetry
of the Minkowski boundary of $AdS$ was introduced in \cite{9807115}, \cite{Kallosh2}, \cite{PST} for the study of brane models on the $AdS\times S$
backgrounds.}. To make the exposition more self-contained
we start by reviewing the $OSp(4|8)/(SO(1,3)\times SO(7))$
supercoset membrane action and its dimensional reduction to the
superstring action on $AdS_4\times\mathbb{CP}^3$. The details of the notation, spinor and isometry algebras are taken to the Appendixes.

\section{The supermembrane on $AdS_4\times S^7$ and its reduction to the $D=10$ IIA superstring on $AdS_4\times\mathbb{CP}^3$}

The $D=11$ supermembrane action on the $AdS_4\times S^7$ background \cite{Plefka98} is given by
\begin{equation}\label{membrane}
S=-\int\limits_{V} d^3\xi\sqrt{-g^{(3)}}+S_{WZ},
\end{equation}
where $g^{(3)}$ is the determinant of the induced world-volume metric
\begin{equation}
g^{(3)}_{\underline i\underline j}=E^{\hat m}_{\underline
i}E_{\underline j\hat m}
\end{equation}
and the Wess-Zumino (WZ) term
\begin{equation}
S_{WZ}=s\int\limits_{\mathcal M_4} H_{(4)}
\end{equation}
is presented as the integral of the closed 4-form
\begin{equation}
H_{(4)}=\frac{i}{2}F^{\hat\alpha}\wedge\mathfrak{g}^{\hat m\hat
n}{}_{\hat\alpha}{}^{\hat\beta}F_{\hat\beta} \wedge E_{\hat
m}\wedge E_{\hat n}+\varepsilon_{m'n'k'l'}E^{m'}\wedge
E^{n'}\wedge E^{k'}\wedge E^{l'}
\end{equation}
over the 4-dimensional auxiliary hypersurface $\mathcal M_4$, whose boundary coincides with the supermembrane world volume $V$.
The first term has the same structure as in the flat superspace
\cite{BSTown}, while the second term is the contribution of the nonzero bosonic
4-form of the background. The $D=11$ supervielbein bosonic components $E^{\hat
m}=(E^{m'},E^{I'})=(G^{0'm'},\Omega^{8I'})$ consist of the $AdS_4$
and $S^7$ vielbeins $G^{0'm'}$ and $\Omega^{8I'}$ that are the Cartan forms
corresponding to the $so(2,3)/so(1,3)$ and $so(8)/so(7)$
 generators $M_{0'm'}$ and $V^{8'I'}$ respectively. Together with the
fermionic 1-forms $F^{\hat\alpha}\equiv F^{\alpha A'}$ associated
with the $osp(4|8)$ fermionic generators $O_{\hat\alpha}\equiv
O_{\alpha A'}$ they satisfy the $osp(4|8)$ Maurer-Cartan (MC)
equations
\begin{equation}\label{MCmemb}
\begin{array}{rl}
dG^{0'm'}&=2G^{m'}{}_{n'}\wedge G^{0'n'}+\frac{i}{4}F^{\alpha A'}\wedge C_{A'B'}\Gamma^{m'}_{\alpha\beta}F^{\beta B'},\\
d\Omega^{8I'}&=2\Omega^{I'J'}\wedge\Omega^{8J'}-\frac{i}{4}F^{\alpha A'}\wedge\Gamma^5_{\alpha\beta}\gamma^{I'}_{A'B'}F^{\beta B'},\\
dF^{\alpha A'}&=-\frac12G^{m'n'}\wedge F^{\beta
A'}\Gamma_{m'n'\beta}{}^\alpha-G^{0'm'}\wedge F^{\beta
A'}\Gamma_{m'\beta}{}^\gamma\Gamma^5{}_\gamma{}^\alpha\\
&+\frac12\Omega^{I'J'}\wedge F^{\alpha
B'}\gamma^{I'J'}{}_{B'}{}^{A'}-\frac12\Omega^{8I'}\wedge F^{\alpha
B'}\gamma^{I'}{}_{B'}{}^{A'}
\end{array}
\end{equation}
that can be derived by taking into account Eq.(\ref{cf}) below and the $osp(4|8)$ (anti)commutation relations in the form (\ref{so8in7}), (\ref{antic}), (\ref{so8fermitransin7}), (\ref{so23fermitransin4}) given in Appendix B. These MC equations are used to show the closeness of $H_{(4)}$.

To verify the $\kappa-$invariance of the membrane action (\ref{membrane}) consider the following variation of the fermionic vielbein components
\begin{equation}\label{kappavar}
F_{\hat\alpha}(\delta_\kappa)=\Pi_{\hat\alpha}{}^{\hat\beta}\kappa_{\hat\beta}(\xi),
\end{equation}
where the matrix $\Pi_{\hat\alpha}{}^{\hat\beta}$ has the form
\begin{equation}
\Pi_{\hat\alpha}{}^{\hat\beta}=\delta^{\hat\beta}_{\hat\alpha}+\frac{q}{\sqrt{-g^{(3)}}}\varepsilon^{\underline i\underline j\underline k}E^{\hat m_1}_{\underline i}E^{\hat m_2}_{\underline j}E^{\hat m_3}_{\underline k}\mathfrak{g}_{\hat m_1\hat m_2\hat m_3\hat\alpha}{}^{\hat\beta},
\end{equation}
accompanied by $E^{\hat m}(\delta_\kappa)=0$. Using that $dH_{(4)}=0$ the variation of $H_{(4)}$ reduces to
\begin{equation}\label{Hvar}
\delta_\kappa H_{(4)}=id\left(F^{\hat\alpha}\mathfrak{g}_{\hat m\hat n\hat\alpha}{}^{\hat\beta}F_{\hat\beta}(\delta_\kappa)\wedge E^{\hat m}\wedge E^{\hat n}\right).
\end{equation}
Substituting (\ref{kappavar}) into (\ref{Hvar}) and performing the $\gamma-$matrix
rearrangements using the formulae adduced in \cite{TFSusy} yields the expression for the
Wess-Zumino 4-form variation
\begin{equation}
\delta_\kappa H_{(4)}=id\left(F^{\hat\alpha}\mathfrak{g}_{\hat m\hat
n\hat\alpha}{}^{\hat\beta}\kappa_{\hat\beta}\wedge E^{\hat
m}\wedge E^{\hat n}-\frac{6q}{\sqrt{-g^{(3)}}}\varepsilon^{\underline
i\underline j\underline k}\varepsilon^{\underline i'\underline
j'\underline k'}(E_{\underline i}\cdot E_{\underline
i'})(E_{\underline j}\cdot E_{\underline j'})E_{\hat m\underline
k'}F^{\hat\alpha}_{\underline k}\mathfrak{g}^{\hat
m}{}_{\hat\alpha}{}^{\hat\beta}\kappa^{\hat\beta}\right).
\end{equation}
It has to be canceled by the $\kappa-$variation of the kinetic
term of (\ref{membrane}) that is defined by the variation of the
induced metric determinant $g^{(3)}$. Taking into account that the
$\kappa-$variation of the bosonic vielbein reads
\begin{equation}
\delta_\kappa E^{\hat m}=-\frac{i}{2}F^{\hat\alpha}\mathfrak{g}^{\hat m}{}_{\hat\alpha}{}^{\hat\beta}F_{\hat\beta}(\delta_\kappa)
\end{equation}
one obtains
\begin{equation}
\delta_\kappa\sqrt{-g^{(3)}}=\frac{i}{4\sqrt{-g^{(3)}}}\varepsilon^{\underline
i\underline j\underline k}\varepsilon^{\underline i'\underline
j'\underline k'}(E_{\underline i}\cdot E_{\underline
i'})(E_{\underline j}\cdot E_{\underline j'})E^{\hat
m}_{\underline k'}F^{\hat\alpha}_{\underline k}\mathfrak{g}_{\hat
m\hat\alpha}{}^{\hat\beta}\kappa_{\hat\beta}-\frac{3iq}{2}F^{\hat\alpha}\mathfrak{g}^{\hat
m\hat n}{}_{\hat\alpha}{}^{\hat\beta}\kappa_{\hat\beta}\wedge
E_{\hat m}\wedge E_{\hat n}.
\end{equation}
The $\kappa-$invariance condition of the action (\ref{membrane})
then fixes the values of the numerical coefficients
$s=\pm\frac14$, $q=\mp\frac16$ \footnote{Such choice of the value
of $q$ makes the matrix $\Pi_{\hat\alpha}{}^{\hat\beta}$ the
projector that eliminates half components of the transformation
parameter $\kappa^{\hat\alpha}(\xi)$.} so that the final form of
the WZ term is given by
\begin{equation}
S_{WZ}=\pm\frac14\int\limits_{\mathcal M_4}\left(\frac{i}{2}F^{\hat\alpha}\wedge\mathfrak{g}^{\hat m\hat
n}{}_{\hat\alpha}{}^{\hat\beta}F_{\hat\beta} \wedge E_{\hat
m}\wedge E_{\hat n}+\varepsilon_{m'n'k'l'}E^{m'}\wedge
E^{n'}\wedge E^{k'}\wedge E^{l'}\right).
\end{equation}

Dimensional reduction of the $D=11$ supermembrane action to the
$D=10$ Type IIA superstring was described for general
superbackground in \cite{DHIS}. The vielbeins are chosen not to
depend on the reduction direction coordinate $y\in[0,2\pi)$.
However, when the bosonic components of the supervielbein receive
contribution $E^{\hat m'}_y$ proportional to the differential $dy$
of the $U(1)$ coordinate\footnote{As is the case for the
$OSp(4|8)/(SO(1,3)\times SO(7))$ supercoset elements considered in
\cite{GSWnew} and here.} the local Lorentz rotation in the tangent
space has to be performed
\begin{equation}
E^{\hat m}\rightarrow L^{\hat m}{}_{\hat n}E^{\hat n},\quad
F^{\hat\alpha}\rightarrow
L^{\hat\alpha}{}_{\hat\beta}F^{\hat\beta},\quad L^{\hat m}{}_{\hat
n}\in SO(1,10),\ L^{\hat\alpha}{}_{\hat\beta}\in Spin(1,10)
\end{equation}
with the parameters determined by $E^{\hat m'}_y$ to bring the
supervielbein bosonic components to the Kaluza-Klein ansatz form
\begin{equation}\label{kkanzatz}
(LE)^{\hat m'}=\mathrm E^{\hat m'},\quad (LE)^{11}=\mathrm E^{11}=\Phi(dy+A),
\end{equation}
where $\mathrm E^{\hat m'}$ is the bosonic part of the $D=10$
supervielbein that does not depend on $y$ and $dy$, the $D=10$ IIA
superfield $\Phi=e^{2\phi/3}$ is related to the dilaton $\phi$,
and $A$ is the RR 1-form potential. For the fermionic components
of the $D=11$ supervielbein the reduction ansatz is
\begin{equation}\label{kkfermviel}
(LF)^{\hat\alpha}=E^{\hat\alpha}+e^{-2\phi/3}\chi^{\hat\alpha}\mathrm E^{11}:\quad
E^{\hat\alpha}=(Lf)^{\hat\alpha}-(LF_y)^{\hat\alpha}A,\
\chi^{\alpha}=(LF_y)^{\hat\alpha},
\end{equation}
where $E^{\hat\alpha}$ are the $D=10$ supervielbein fermionic
components, $\chi^{\hat\alpha}$ is the dilatino superfield and there has been
separated the $dy$-dependent contribution to the $D=11$ supervielbein fermionic components
\begin{equation}\label{fermviel}
F^{\hat\alpha}=f^{\hat\alpha}+dyF^{\hat\alpha}_y.
\end{equation}

Upon identification of $y$ with the world-volume compact direction coordinate the kinetic term of the membrane action (\ref{membrane}) reduces to the Nambu-Goto string action in the Kaluza-Klein frame
\begin{equation}
\int\limits_V d^3\xi\sqrt{-g^{(3)}}\rightarrow\int\limits_\Sigma
d\tau d\sigma\Phi\sqrt{-g^{(2)}},
\end{equation}
where $g^{(2)}=\det g^{(2)}_{ij}$ is the determinant of the
induced world-sheet metric
\begin{equation}
g^{(2)}_{ij}=\mathrm E^{\hat m'}_i\mathrm E_{\hat m'j},
\end{equation}
while the membrane WZ term reduces to the integral of the NS-NS 3-form over the auxiliary 3-dimensional hypersurface $\mathcal M_3$, whose boundary coincides with the superstring world-sheet $\Sigma$
\begin{equation}\label{wzred}
\int\limits_{\mathcal M_4} H_{(4)}\rightarrow \int\limits_{\mathcal M_3}H_{(3)}.
\end{equation}

To specificate the above discussion let us describe the
transformation of the $so(8)$ Cartan forms to the
$\mathbb{CP}^3\times U(1)$-basis that exhibits the Hopf fibration
realization of $S^7$ necessary to perform the double dimensional
reduction and also of the $sp(4)$ Cartan forms to the $D=3$
conformal basis. For the $OSp(4|8)/(SO(1,3)\times SO(7)$ supercoset element $\mathscr G$ the $SO(1,3)\times SO(7)$ covariant decomposition
of the Cartan forms reads
\begin{equation}\label{cf}
\mathscr G^{-1}d\mathscr G=G^{m'n'}M_{m'n'}+G^{0'm'}M_{0'm'}+\Omega^{I'J'}V^{I'J'}+\Omega^{8I'}V^{8I'}+F^{\alpha
A'}O_{\alpha A'}.
\end{equation}
The first two terms correspond to the $D=3$ conformal sector
and after the transformation of the generators to the conformal basis, as described in Appendix B, are brought to the form
\begin{equation}
G^{m'n'}M_{m'n'}+G^{0'm'}M_{0'm'}=G^{mn}M_{mn}+\Delta D+\omega^m P_m+c^m K_m,
\end{equation}
where
\begin{equation}
\omega^m=G^{0'm}-G^{3m},\quad c^m=G^{0'm}+G^{3m},\quad \Delta=-G^{0'3}.
\end{equation}
The $AdS_4$ vielbein components, used in constructing the superstring action, in the conformal basis are expressed as
\begin{equation}\label{adsvielbein}
G^{0'm}=\frac12(\omega^m+c^m),\quad G^{0'3}=-\Delta.
\end{equation}

The third and the fourth terms in (\ref{cf}) correspond to the $so(8)$ generators and 1-forms and further decompose as
\begin{equation}\label{cfso8}
\Omega^{I'J'}V^{I'J'}+\Omega^{8I'}V^{8I'}=\Omega^{IJ}V^{IJ}+\Omega^{78}V^{78}+\Omega^{8I}V^{8I}+2\Omega^{7I}V^{7I}.
\end{equation}
To perform the dimensional reduction of the $AdS_4\times S^7$ membrane action one is required to change the generator basis to that corresponding to
the Hopf fibration realization of the 7-sphere (see Appendix B).
Then the two last terms in (\ref{cfso8}) corresponding to the $so(8)/(so(2)\times
so(6))$ coset generators acquire the form
\begin{equation}
\Omega^{8I}V^{8I}+2\Omega^{7I}V^{7I}=\Omega_a T^a+\Omega^aT_a+\tilde\Omega_a\tilde T^a+\tilde\Omega^a\tilde T_a,
\end{equation}
where
\begin{equation}
\begin{array}{c}
\Omega_a=\Omega^7{}_{4a}-\frac{i}{2}\Omega^8{}_{4a},\quad\Omega^a=-\Omega^{74a}-\frac{i}{2}\Omega^{84a},\\
\tilde\Omega_a=-\Omega^7{}_{4a}-\frac{i}{2}\Omega^8{}_{4a},\quad\tilde\Omega^a=\Omega^{74a}-\frac{i}{2}\Omega^{84a}
\end{array}
\end{equation}
and $\Omega^{7(8)}{}_{4a}=\Omega^{7(8)I}\rho^I_{4a}$, $\Omega^{7(8)4a}=\Omega^{7(8)I}\tilde\rho^{I4a}$. The 1-forms
\begin{equation}
\Omega^8{}_{4a}=i(\Omega_a+\tilde\Omega_a),\quad\Omega^{84a}=i(\Omega^a+\tilde\Omega^a)
\end{equation}
will be used in constructing the superstring action.
Changing the basis in the $so(6)\oplus so(2)$ sector that is described by the first two terms in (\ref{cfso8}) yields
\begin{equation}\label{cfso6so2}
\Omega^{IJ}V^{IJ}+\Omega^{78}V^{78}=\tilde\Omega_a{}^b\tilde
V_b{}^a+\tilde\Omega_b{}^b\tilde
V_a{}^a+\Omega_a{}^4V_4{}^a+\Omega_4{}^aV_a{}^4+hH,
\end{equation}
where
\begin{equation}
\begin{array}{c}
\tilde\Omega_a{}^b=\Omega_a{}^b-\delta^b_a\Omega_c{}^c+\delta^b_ah,\quad\tilde\Omega_b{}^b=-2\Omega_b{}^b+3h,\\
\Omega^{78}=-\tilde\Omega_a{}^a-h,
\end{array}
\end{equation}
and the $so(6)$ Cartan forms have been decomposed into the $u(3)$ $\Omega_a{}^b$ and $su(4)/u(3)$ $\Omega_a{}^4$, $\Omega_4{}^a$ ones as
\begin{equation}
\Omega_A{}^B=\frac{i}{2}\Omega^{IJ}\rho^{IJ}{}_A{}^B=\left(
\begin{array}{cc}
\Omega_a{}^b &\Omega_a{}^4\\
\Omega_4{}^b &\Omega_4{}^4
\end{array}\right),\quad
\Omega_4{}^4=-\Omega_a{}^a.
\end{equation}
The 1-form $h$ in (\ref{cfso6so2}) corresponds to the fiber direction of $\mathbb{CP}^3\times U(1)$.

The fermionic sector of (\ref{cf}) decomposes as follows
\begin{equation}
\begin{array}{c}
F^{\alpha A'}O_{\alpha A'}=\bar F^{\alpha A}\bar O_{\alpha A}+F^\alpha_AO^A_\alpha\\
=\omega^\mu_A Q^A_\mu+\bar\omega^{\mu A}\bar Q_{\mu A}+\chi_{\mu A}S^{\mu A}+\bar\chi^A_\mu\bar S^\mu_A,
\end{array}
\end{equation}
where the generators have been split into the supersymmetry and superconformal ones (see Eq.(\ref{susygen})) accompanied by the corresponding
splitting of the Cartan forms
\begin{equation}\label{defsconff}
\bar F^{\alpha A}=\sqrt{2}\left(
\begin{array}{c}
\bar\omega^{\mu A}\\
\bar\chi^A_\mu\\
\end{array}\right),\quad
F^\alpha_A=\sqrt{2}\left(
\begin{array}{c}
\omega^\mu_A\\
\chi_{\mu A}\\
\end{array}\right).
\end{equation}

The MC equations (\ref{MCmemb}) in the novel basis
acquire the form
\begin{equation}
\begin{array}{rcl}
d\omega^m&=&2G^m{}_n\wedge\omega^n+2\Delta\wedge\omega^m-2i\omega^\mu_A\sigma^m_{\mu\nu}\bar\omega^{\nu A},\\
dc^m&=&2G^m{}_n\wedge c^n-2\Delta\wedge c^m-2i\chi_{\mu A}\tilde\sigma^{m\mu\nu}\bar\chi^A_\nu,\\
d\Delta&=&\omega^m\wedge
c_m+i\omega^\mu_A\wedge\bar\chi^A_\mu+i\bar\omega^{\mu
A}\wedge\chi_{\mu A}
\end{array}
\end{equation}
for the Cartan forms from the $AdS$ sector,
\begin{equation}
\begin{array}{rcl}
d\Omega_a&=&-i\tilde\Omega_a{}^b\wedge\Omega_b-i\tilde\Omega_b{}^b\wedge\Omega_a-i\varepsilon_{abc}\Omega_4{}^b\wedge\tilde\Omega^c+2\varepsilon_{abc}\bar\omega^{\mu b}\wedge\bar\chi^c_\mu,\\
d\Omega^a&=&-i\Omega^b\wedge\tilde\Omega_b{}^a-i\Omega^a\wedge\tilde\Omega_b{}^b+i\varepsilon^{abc}\Omega_b{}^4\wedge\tilde\Omega_c-2\varepsilon^{abc}\omega^\mu_b\wedge\chi_{\mu c},\\
d\tilde\Omega_a&=&-i\tilde\Omega_a{}^b\wedge\tilde\Omega_b+i\tilde\Omega_b{}^b\wedge\tilde\Omega_a+2ih\wedge\tilde\Omega_a-i\varepsilon_{abc}\Omega_4{}^b\wedge\Omega^c-2\omega^{\mu}_{a}\wedge\chi_{\mu 4}+2\omega^\mu_4\wedge\chi_{\mu a},\\
d\tilde\Omega^a&=&-i\tilde\Omega^b\wedge\tilde\Omega_b{}^a+i\tilde\Omega^a\wedge\tilde\Omega_b{}^b-2ih\wedge\tilde\Omega^a+i\varepsilon^{abc}\Omega_b{}^4\wedge\Omega_c+2\bar\omega^{\mu a}\wedge\bar\chi_{\mu 4}-2\bar\omega^{\mu 4}\wedge\bar\chi^a_\mu,\\
d\tilde\Omega_a{}^a&=&-i\Omega_a\wedge\Omega^a+\frac{i}{2}\tilde\Omega_a\wedge\tilde\Omega^a+\frac{i}{2}\Omega_a{}^4\wedge\Omega_4{}^a-\omega^\mu_a\wedge\bar\chi^a_\mu+\bar\omega^{\mu a}\wedge\chi_{\mu a},\\
dh&=&\frac{i}{2}\tilde\Omega_a\wedge\tilde\Omega^a+\frac{i}{2}\Omega_4{}^a\wedge\Omega_a{}^4-\omega^\mu_4\wedge\bar\chi^4_\mu+\bar\omega^{\mu
4}\wedge\chi_{\mu 4}
\end{array}
\end{equation}
for the Cartan forms from the $so(8)$ sector and
\begin{equation}
\begin{array}{rcl}
d\omega^\mu_a&=&\Delta\wedge\omega^\mu_a+\frac12G^{mn}\wedge\omega^\nu_a\sigma_{mn\nu}{}^\mu-\omega^m\wedge\tilde\sigma_m^{\mu\nu}\chi_{\nu a}-i\tilde\Omega_a{}^b\wedge\omega^\mu_b+i\tilde\Omega_b{}^b\wedge\omega^\mu_a\\
&-&i\Omega_a{}^4\wedge\omega^\mu_4+i\tilde\Omega_a\wedge\bar\omega^{\mu 4}-i\varepsilon_{abc}\Omega^b\wedge\bar\omega^{\mu c},\\
d\omega^\mu_4&=&\Delta\wedge\omega^\mu_4+\frac12G^{mn}\wedge\omega^\nu_4\sigma_{mn\nu}{}^\mu-\omega^m\wedge\tilde\sigma_m^{\mu\nu}\chi_{\nu 4}+2ih\wedge\omega^\mu_4-i\Omega_4{}^a\wedge\omega^\mu_a-i\tilde\Omega_a\wedge\bar\omega^{\mu a},\\
d\chi_{\mu a}&=&-\Delta\wedge\chi_{\mu a}-\frac12G^{mn}\wedge\sigma_{mn\mu}{}^\nu\chi_{\nu a}+c^m\wedge\sigma_{m\mu\nu}\omega^\nu_a-i\tilde\Omega_a{}^b\wedge\chi_{\mu b}+i\tilde\Omega_b{}^b\wedge\chi_{\mu a}\\
&-&i\Omega_a{}^4\wedge\chi_{\mu 4}+i\tilde\Omega_a\wedge\bar\chi^4_\mu-i\varepsilon_{abc}\Omega^b\wedge\bar\chi^c_\mu,\\
d\chi_{\mu 4}&=&-\Delta\wedge\chi_{\mu
4}-\frac12G^{mn}\wedge\sigma_{mn\mu}{}^\nu\chi_{\nu
4}+c^m\wedge\sigma_{m\mu\nu}\omega^\nu_4+2ih\wedge\chi_{\mu
4}-i\Omega_4{}^a\wedge\chi_{\mu
a}-i\tilde\Omega_a\wedge\bar\chi^a_\mu
\end{array}
\end{equation}
and c.c. for the fermionic 1-forms.

In what follows we shall specialize to the $OSp(4|8)/(SO(1,3)\times SO(7))$ representative in the form of the "dressed" $OSp(4|6)/(SO(1,3)\times U(3))$ supercoset element
\begin{equation}\label{48cosetel}
\mathscr G=\mathscr G_{OSp(4|6)/(SO(1,3)\times
U(3))}e^{yH}e^{\theta^\mu_4Q^4_\mu+\bar\theta^{\mu 4}\bar Q_{\mu
4}}e^{\eta_{\mu 4}S^{\mu 4}+\bar\eta^4_\mu\bar S^\mu_4},
\end{equation}
where
\begin{equation}\label{46cosetel}
\mathscr G_{OSp(4|6)/(SO(1,3)\times U(3))}=e^{x^mP_m+\theta^\mu_aQ^a_\mu+\bar\theta^{\mu a}\bar Q_{\mu a}}e^{\eta_{\mu a}S^{\mu a}+\bar\eta^a_\mu\bar S^{\mu}_a}e^{z^aT_a+\bar z_aT^a}e^{\varphi D}
\end{equation}
is the $OSp(4|6)/(SO(1,3)\times U(3))$ supercoset element considered in \cite{0811}. It is
parametrized by the $D=3$ $\mathcal N=6$ super-Poincare coordinates
$x^m,\theta^\mu_a,\bar\theta^{\mu a}$ supplemented by the
coordinates $\eta_{\mu a},\bar\eta^a_\mu$ associated with the
superconformal generators, the coordinates $z^a,\bar z_a$
parametrizing $\mathbb{CP}^3$ and the coordinate $\varphi$ related to the radial
direction of $AdS_4$. The expressions for the Cartan forms corresponding
to the supercoset element (\ref{46cosetel}) have been derived in \cite{0811}.
The form of the supercoset element (\ref{48cosetel}) is governed
by the requirement of the absence of the vielbein dependence on
the reduction direction coordinate $y$ that can be satisfied by
placing $e^{yH}$ to the left from the factors corresponding to
the broken supersymmetries of the background, whenever the differential acts from
the right. The differentials of the factors corresponding to the broken supersymmetries give contributions to the $sp(4)$
 Cartan forms, the forms associated with the broken
supersymmetries and the $U(1)$ direction. After the commutation of
$\mathscr G^{-1}_{OSp(4|6)/(SO(1,3)\times U(3))}d\mathscr G_{OSp(4|6)/(SO(1,3)\times U(3))}$ with the factors
corresponding to the broken supersymmetries the $osp(4|6)$ Cartan
forms $\mathcal C$ become $\mathcal N=4$ superfields $\mathcal
C(\theta_4,\bar\theta^4,\eta_4,\bar\eta^4)$. There also appear the
contributions proportional to the broken supersymmetries generators and
$\tilde T_a$, $\tilde T^a$, $V_a{}^4$, $V_4{}^a$, $H$ generators.

\section{$AdS_4\times\mathbb{CP}^3$ superstring in the light-cone gauge}

The fermionic light-cone gauge condition we consider
\begin{equation}\label{lcfermi}
\theta^2_A=\bar\theta^{2 A}=\eta_{1 A}=\bar\eta_{1}^{A}=0
\end{equation}
is characterized by setting to zero the coordinates associated
with the generators $Q^A_2$, $\bar Q_{2 A}$, $S^{1A}$, $\bar
S^1_A$ that have negative charge w.r.t. the $so(1,1)$ generator
$M^{+-}\equiv2M^{02}$ from the Lorentz group acting on the
Minkowski boundary of $AdS_4$
\begin{equation}
[M^{+-},Q^A_2]=-Q^A_2,\quad [M^{+-},\bar Q_{2A}]=-\bar Q_{2A},\quad [M^{+-},S^{1A}]=-S^{1A},\quad [M^{+-},\bar S^1_A]=-\bar S^1_A.
\end{equation}
Other fermionic generators have positive $so(1,1)$ charge
\begin{equation}
[M^{+-},Q^A_1]=Q^A_1,\quad [M^{+-},\bar Q_{1A}]=\bar Q_{1A},\quad [M^{+-},S^{2A}]=S^{2A},\quad [M^{+-},\bar S^2_A]=\bar S^2_A.
\end{equation}
Respective coordinates can be denoted as follows
\begin{equation}
\theta^1_A\equiv\theta^-_A\equiv\theta_A,\quad\bar\theta^{1A}\equiv\bar\theta^{-A}\equiv\bar\theta^A,\quad
\eta^{1}_{A}\equiv\eta^{-}_{A}\equiv\eta_A,\quad\bar\eta^{1A}\equiv\bar\eta^{-A}\equiv\bar\eta^A
\end{equation}
and become the physical degrees of freedom of the superstring in the gauge (\ref{lcfermi}).

In the light-cone gauge the constituents of the $AdS_4$ vielbein (\ref{adsvielbein}) reduce to
\begin{equation}\label{omega}
\begin{array}{c}
\omega^m=e^{-2\varphi}(dx^m-id\theta_a\sigma^m\bar\theta^a+i\theta_a\sigma^md\bar\theta^a)-id\theta_4\sigma^m\bar\theta^4+i\theta_4\sigma^md\bar\theta^4+dy\omega^m_y,\quad\omega^m_y=4\theta_4\sigma^m\bar\theta^4,\\
c^m=-ie^{2\varphi}(d\eta_a\tilde\sigma^m\bar\eta^a-\eta_a\tilde\sigma^md\bar\eta^a)-
id\eta_4\tilde\sigma^m\bar\eta^4+i\eta_4\tilde\sigma^md\bar\eta^4+dyc^m_y,\quad c^m_y=4\eta_4\tilde\sigma^m\bar\eta^4,\\
\Delta=d\varphi.
\end{array}
\end{equation}
The $\mathbb{CP}^3\times U(1)$ vielbein components acquire the form
\begin{equation}\label{omega84a}
\Omega^8{}_{4a}=i(\Omega_a+\varepsilon_{abc}\hat{\bar\eta}{}^b\hat{\bar\eta}{}^cdx^+-2e^{-\varphi}\hat\eta_a\eta_4dx^+),\quad\Omega^{84a}=i(\Omega^a-\varepsilon^{abc}\hat\eta_b\hat\eta_cdx^++2e^{-\varphi}\hat{\bar\eta}{}^a\bar\eta^4dx^+),
\end{equation}
\begin{equation}
\Omega^{87}=h+\tilde\Omega_a{}^a,
\end{equation}
where
\begin{equation}
h=dy-e^{-2\varphi}\eta_4\bar\eta^4dx^+,
\end{equation}
\begin{equation}
\hat\eta_a=T_a{}^b\eta_b+T_{ab}\bar\eta^b,\quad\hat{\bar\eta}{}^a=T^a{}_b\bar\eta^b+T^{ab}\eta_b
\end{equation}
and the matrix $T_{\hat a}{}^{\hat b}$ has been defined in Eq.(64) of \cite{0811}.
The Cartan forms $\Omega_a$, $\Omega^a$ and $\tilde\Omega_a{}^a$ in the light-cone gauge become purely bosonic quantities equal to
$\Omega_{\mathbf ba}{}^4$, $\Omega_{\mathbf b4}{}^a$ and $\Omega_{\mathbf ba}{}^a$ respectively given by Eq.(65) of \cite{0811}. Then the $S^7$ part of the target-space metric acquires the Kaluza-Klein form
\begin{equation}
ds^2_{S^7}=\Omega^{8I'}\Omega^{8I'}=\Omega^{87}\Omega^{87}+ds^2_{CP^3}=(dy+a)^2+ds^2_{CP^3},
\end{equation}
where
\begin{equation}
a(d)=\tilde\Omega_a{}^a-e^{-2\varphi}\eta_4\bar\eta^4dx^+
\end{equation}
and
\begin{equation}
ds^2_{CP^3}=-\Omega^{8}{}_{4a}\Omega^{84a}.
\end{equation}

In analogy with (\ref{fermviel}) in the Cartan forms (\ref{defsconff}) associated with the supersymmetry and superconformal generators it is also
possible to single out the terms proportional to the reduction direction coordinate differential $dy$
\begin{equation}
\omega^\mu_A=\tilde\omega^\mu_A+dy\omega^\mu_{yA},\quad\bar\omega^{\mu
A}=\bar{\tilde\omega}{}^{\mu A}+dy\bar\omega^{\mu A}_y,\quad
\chi_{\mu A}=\tilde\chi_{\mu A}+dy\chi_{y\mu
A},\quad\bar\chi^A_\mu=\bar{\tilde\chi}{}^A_\mu+dy\bar{\chi}{}^A_{y\mu}.
\end{equation}
In the light-cone gauge the $dy-$independent components are given by
\begin{equation}\label{omega-a}
\tilde\omega^\mu_a=e^{-\varphi}\left(
\begin{array}{c}
\hat d\theta_a+dx^1\hat\eta_a\\
dx^+\hat\eta_a
\end{array}\right),\quad
\bar{\tilde\omega}{}^{\mu a}=e^{-\varphi}\left(
\begin{array}{c}
\hat d\bar\theta^a+dx^1\hat{\bar\eta}{}^a\\
dx^+\hat{\bar\eta}{}^a
\end{array}\right);
\end{equation}
\begin{equation}\label{chi-a}
\tilde\chi_{\mu a}=\left(
\begin{array}{c}
0\\
e^{\varphi}\hat d\eta_a
\end{array}\right),\quad
\bar{\tilde\chi}{}^a_\mu=\left(
\begin{array}{c}
0\\
e^{\varphi}\hat d\bar\eta^a
\end{array}\right),
\end{equation}
where $\hat d\theta_a(\hat d\eta_a)=T_a{}^bd\theta_b(d\eta_b)+T_{ab}d\bar\theta^b(d\bar\eta^b)$, $\hat
d\bar\theta^a(\hat d\bar\eta^a)=T^a{}_bd\bar\theta^b(d\bar\eta^b)+T^{ab}d\theta_b(d\eta_b)$, and
\begin{equation}\label{omega-4}
\tilde\omega^\mu_4=\left(
\begin{array}{c}
d\theta_4+d\varphi\theta_4+e^{-2\varphi}dx^1\eta_4\\
e^{-2\varphi}dx^+\eta_4
\end{array}\right),\quad
\bar{\tilde\omega}{}^{\mu 4}=\left(
\begin{array}{c}
d\bar\theta^4+d\varphi\bar\theta^4+e^{-2\varphi}dx^1\bar\eta^4\\
e^{-2\varphi}dx^+\bar\eta^4
\end{array}\right);
\end{equation}
\begin{equation}
\tilde\chi_{\mu 4}=\left(
\begin{array}{c}
0\\
d\eta_4-d\varphi\eta_4
\end{array}\right),\quad
\bar{\tilde\chi}{}^4_\mu=\left(
\begin{array}{c}
0\\
d\bar\eta^4-d\varphi\bar\eta^4
\end{array}\right).
\end{equation}
The components of $F^{\hat\alpha}_y$ acquire the form
\begin{equation}\label{omega-y}
\omega^\mu_{y4}=\left(
\begin{array}{c}
2i\theta_4\\
0
\end{array}\right),\quad
\bar\omega^{\mu 4}_y=\left(
\begin{array}{c}
-2i\bar\theta^4\\
0
\end{array}\right);
\end{equation}
\begin{equation}\label{chiy}
\chi_{y\mu 4}=\left(
\begin{array}{c}
0\\
2i\eta_4
\end{array}\right),\quad
\bar\chi^4_{y\mu}=\left(
\begin{array}{c}
0\\
-2i\bar\eta^4
\end{array}\right).
\end{equation}

For the $OSp(4|8)/(SO(1,3)\times SO(7))$ supercoset element (\ref{48cosetel}) the
consequence of the absence of the vielbein dependence on the
reduction direction coordinate $y$ is that the $AdS_4$ vielbein
components (\ref{omega}) acquire the contributions proportional to
$dy$. To remove them, i.e. to bring the bosonic components of the supervielbein to the
Kaluza-Klein ansatz form (\ref{kkanzatz}), the local Lorentz
rotation in the tangent space needs to be performed. Since the
$\mathbb{CP}^3$ vielbein components (\ref{omega84a}) do not
contain the contributions proportional to $dy$, the necessary
frame rotation $L$ involves only tangent to the $AdS_4$ directions and
the one tangent to the U(1)-fiber direction on $S^7$ labeled by
"11" in the $11d$ context
\begin{equation}
\left(\begin{array}{c}
\mathrm E^{m'}\\
\mathrm E^{11}
\end{array}\right)=
L\left(\begin{array}{c}
G^{0'm'}\\
\Omega^{87}
\end{array}\right).
\end{equation}
The entries of the matrix $L$
\begin{equation}
L=\left(\begin{array}{cc}
L^{m'}{}_{n'}& L^{m'}{}_{7}\\
L^7{}_{m'}&L^7{}_7
\end{array}\right)\in SO(1,4)
\end{equation}
in the light-cone gauge are given by
\begin{equation}
L^{m'}{}_{n'}=\delta^{m'}_{n'}-\frac12G^{0'm'}_yG^{0'}_{yn'},\quad
L^{m'}{}_7=-G^{0'm'}_y,\quad L^{7}{}_{m'}=G^{0'}_{ym'},\quad
L^7{}_7=1,
\end{equation}
where
\begin{equation}
G^{0'm'}_y=\left(\frac12(\omega^m_y+c^m_y),0\right)=2\Theta(1,0,-1,0),\quad \Theta=\theta_4\bar\theta^4+\eta_4\bar\eta^4
\end{equation}
is the $D=1+3$ light-like vector.  Corresponding Lorentz rotation acting on the
supervielbein fermionic components is generated by the matrix
\begin{equation}
L^{\hat\alpha}{}_{\hat\beta}=\delta^{\hat\alpha}_{\hat\beta}-\frac12G^{0'}_{ym'}\mathfrak{g}^{m'\hat\alpha}{}_{\hat\gamma}\mathfrak{g}^{11\hat\gamma}{}_{\hat\beta}.
\end{equation}
As a result transformed bosonic components of the $D=11$ supervielbein in the light-cone basis equal
\begin{equation}\label{tbv}
\begin{array}{c}
\mathrm E^{+}=\frac12e^{-2\varphi}dx^+,\quad\mathrm E^{-}=\frac12e^{-2\varphi}dx^-+\varpi-2e^{-2\varphi}\Theta^2dx^++4\Theta(\tilde\Omega_a{}^a-e^{-2\varphi}\eta_4\bar\eta^4dx^+),\\
\mathrm E^1=\frac12e^{-2\varphi}dx^1,\quad\mathrm E^3=-d\varphi,\\
\mathrm E^{11}=dy+A(d),\quad A(d)=a(d)-e^{-2\varphi}\Theta dx^+
\end{array}
\end{equation}
where $x^{\pm}=x^2\pm x^0$ and
\begin{equation}
\varpi=ie^{-2\varphi}(d\theta_a\bar\theta^a-\theta_ad\bar\theta^a)+i(d\theta_4\bar\theta^4-\theta_4d\bar\theta^4)+ie^{2\varphi}(d\eta_a\bar\eta^a-\eta_ad\bar\eta^a)+i(d\eta_4\bar\eta^4-\eta_4d\bar\eta^4).
\end{equation}
Note that in the light-cone gauge the superfield
$\Phi=\sqrt{1+G^{0'm'}_yG^{0'}_{ym'}}$ turns to unity for the
chosen normalization. Correspondingly the transformed fermionic
components $(Lf)^{\hat\alpha}$ of the supervielbein read
\begin{equation}\label{tchi-a}
(L\tilde\chi)_{\mu a}=\left(
\begin{array}{c}
0\\
e^{\varphi}\hat d\eta_a+2ie^{-\varphi}\Theta\hat\eta_adx^+
\end{array}\right),\quad
(L\bar{\tilde\chi}){}^a_\mu=\left(
\begin{array}{c}
0\\
e^{\varphi}\hat d\bar\eta^a-2ie^{-\varphi}\Theta\hat{\bar\eta}{}^adx^+
\end{array}\right)
\end{equation}
and
\begin{equation}\label{tchi-4}
(L\tilde\chi)_{\mu 4}=\left(
\begin{array}{c}
0\\
d\eta_4-d\varphi\eta_4+2ie^{-2\varphi}\Theta\eta_4dx^+
\end{array}\right),\quad
(L\bar{\tilde\chi}){}^4_\mu=\left(
\begin{array}{c}
0\\
d\bar\eta^4-d\varphi\bar\eta^4-2ie^{-2\varphi}\Theta\bar\eta^4dx^+
\end{array}\right).
\end{equation}
Other fermionic vielbein components (\ref{omega-a}), (\ref{omega-4}), (\ref{omega-y}), (\ref{chiy}) remain unaffected. Altogether they define the
Kaluza-Klein form of the fermionic vielbein (\ref{kkfermviel}).

Then using (\ref{omega84a}) and (\ref{tbv}) we find the expression for the
induced world-sheet metric in the fermionic light-cone gauge (\ref{lcfermi})
\begin{equation}
g^{(2)}_{ij}=g^{AdS}_{ij}+g^{CP}_{ij},
\end{equation}
where
\begin{equation}
\begin{array}{rl}
g^{AdS}_{ij}=&\frac12(\mathrm E^+_i\mathrm E^-_{j}+\mathrm
E^+_{j}\mathrm E^-_i)+\mathrm E^1_i\mathrm E^1_j+\mathrm
E^3_i\mathrm
E^3_j\\[0.2cm]
=&\frac18e^{-4\varphi}(\partial_i x^+\partial_j x^-+\partial_j
x^+\partial_i x^-)+\frac14e^{-4\varphi}\partial_i x^1\partial_jx^1
+\partial_i\varphi\partial_j\varphi\\[0.2cm]
+&\frac14e^{-2\varphi}[\partial_ix^+(\varpi_j+4\Theta\tilde\Omega_{ja}{}^a)+
\partial_jx^+(\varpi_i+4\Theta\tilde\Omega_{ia}{}^a)]-4e^{-4\varphi}\theta_4\bar\theta^4\eta_4\bar\eta^4\partial_ix^+\partial_jx^+
\end{array}
\end{equation}
and
\begin{equation}
\begin{array}{rl}
g^{CP}_{ij}=&-\frac12(\Omega^{84a}_i\Omega^8_j{}_{4a}+\Omega^{84a}_j\Omega^8_i{}_{4a})\\[0.2cm]
=&\frac12(\Omega_{ia}\Omega^a_j+\Omega_{ja}\Omega^a_i)+e^{-\varphi}[\partial_ix^+(\Omega_{ja}\hat{\bar\eta}{}^a\bar\eta^4
-\Omega^a_j\hat\eta_a\eta_4)+\partial_jx^+(\Omega_{ia}\hat{\bar\eta}{}^a\bar\eta^4-\Omega^a_i\hat\eta_a\eta_4)]\\[0.2cm] 
+&\frac12[\partial_ix^+(\varepsilon_{abc}\Omega^a_j\hat{\bar\eta}{}^b\hat{\bar\eta}{}^c-\varepsilon^{abc}\Omega_{ja}\hat\eta_b\hat\eta_c)+\partial_jx^+(\varepsilon_{abc}\Omega^a_i\hat{\bar\eta}{}^b\hat{\bar\eta}{}^c-\varepsilon^{abc}\Omega_{ia}\hat\eta_b\hat\eta_c)]\\[0.2cm]
+&2[(\hat\eta_a\hat{\bar\eta}{}^a)^2+e^{-\varphi}(\varepsilon_{abc}\hat{\bar\eta}{}^a\hat{\bar\eta}{}^b\hat{\bar\eta}{}^c\bar\eta^4+\varepsilon^{abc}\hat\eta_a\hat\eta_b\hat\eta_c\eta_4)+2e^{-2\varphi}\hat\eta_a\hat{\bar\eta}{}^a\eta_4\bar\eta^4]\partial_ix^+\partial_jx^+.
\end{array}
\end{equation} 
The superstring WZ term is determined by the integral of the NS-NS 3-form
(\ref{wzred})
\begin{equation}
H_{(3)}=\frac{i}{4}(E^{\hat\alpha}\mathfrak{g}^{\hat m'\hat
n'}{}_{\hat\alpha}{}^{\hat\beta}\chi_{\hat\beta}\wedge\mathrm
E_{\hat m'}\wedge\mathrm E_{\hat
n'}+E^{\hat\alpha}\mathfrak{g}^{\hat
m'11}{}_{\hat\alpha}{}^{\hat\beta}\wedge
E_{\hat\beta}\wedge\mathrm E_{\hat
m'})-\varepsilon_{m'n'k'l'}\mathrm E^{m'}\wedge\mathrm
E^{n'}\wedge\mathrm E^{k'}L^{l'}{}_7
\end{equation}
that in the fermionic light-cone gauge (\ref{lcfermi})
can be presented as the total differential of the
2-form
\begin{equation}
\begin{array}{rl}
B_{(2)}=&\frac{1}{2}e^{-4\varphi}(\theta_4\bar\theta^4+\eta_4\bar\eta^4)dx^1\wedge
dx^+\\[0.2cm]
+&\frac14e^{-2\varphi}(d\theta_4\bar\eta^4-d\eta_4\bar\theta^4+\eta_4d\bar\theta^4-\theta_4d\bar\eta^4)\wedge
dx^+\\[0.2cm]
+&ie^{-2\varphi}(\theta_4\bar\eta^4-\eta_4\bar\theta^4)dx^+\wedge\tilde\Omega_{a}{}^a
+ie^{-\varphi}\hat\eta_a\theta_4dx^+\wedge\Omega^a+ie^{-\varphi}\hat{\bar\eta}{}^a\bar\theta^4dx^+\wedge\Omega_a\\[0.2cm]
+&e^{-2\varphi}\hat\eta_a\hat{\bar\eta}{}^adx^1\wedge
dx^++\frac12e^{-2\varphi}(\hat\eta_a\hat d\bar\theta^a+\hat
d\theta_a\hat{\bar\eta}{}^a)\wedge dx^+.
\end{array}
\end{equation}
So the Polyakov representation of the $AdS_4\times\mathbb{CP}^3$ superstring action in the fermionic light-cone gauge (\ref{lcfermi}) becomes
\begin{equation}
S_{l.c.}=-\frac12\int\limits_\Sigma d\tau
d\sigma\sqrt{-\gamma}\gamma^{ij}(g^{AdS}_{ij}+g^{CP}_{ij})\pm\int\limits_\Sigma
d\tau d\sigma B_{(2)},
\end{equation}
where $\gamma_{ij}$ is the auxiliary world-sheet metric. The
bosonic light-cone gauge condition can be fixed in various ways,
in particular, within the Hamiltonian approach \cite{MTlc}, \cite{found}.

\section{Conclusion}

We have obtained the $AdS_4\times\mathbb{CP}^3$ superstring
action in the light-cone gauge, in which both light-like
directions lie on the Minkowski boundary of the $AdS_4$ part of the background. In deriving
it we employed the double dimension reduction of the supermembrane
action on the $AdS_4\times S^7$ background constructed in
\cite{Plefka98} using the supercoset approach. To perform the
dimensional reduction it is necessary to change the generator
basis in the $so(8)$ sector of the $osp(4|8)$ isometry superalgebra of $AdS_4\times S^7$ superbackground to
exhibit the Hopf fibration realization of the 7-sphere. In the
$sp(4)$ sector the generators have been transformed to the $3d$
conformal basis. As a result the fermionic generators and coordinates
naturally split into corresponding to the (un)broken Poincare and conformal
supersymmetries. The kinetic term of the
resultant action includes the contributions up to the 4th order in
the space-time fermions and the WZ term is of the 2nd order similarly to the $AdS_5\times S^5$ superstring case \cite{MTlc}. There
can be taken simplifying limits of the action analogous to those
considered for the $AdS_5\times S^5$ superstring (for review see
\cite{found}). It can also be used to study the string states with zero quantum numbers from the $\mathbb{CP}^3$ sector within the semiclassical approximation \cite{GKP}, \cite{FT}, whose application for the $AdS_4\times\mathbb{CP}^3$ superstring has been initiated in \cite{semiclas}.

\section{Acknowledgements}

The author is obliged to A.A.~Zheltukhin for stimulating discussions.

\appendix
\section{Notation and spinor properties}

We use the following notation
\begin{itemize}
\item for the vector indices
\begin{itemize}
\item in $D=1+10$ dimensions $\hat m,\hat n=0,1,...9,11$;
\item in $D=1+9$ dimensions $\hat m',\hat n'=0,1,...,9$;
\item in $D=8$ dimensions $\underline I,\underline J,\underline K=1,...,8$;
\item in $D=7$ dimensions $I',J',K'=1,...,7$;
\item in $D=6$ dimensions $I,J,K=1,...,6$;
\item in $D=2+3$ dimensions $\underline m,\underline n=0',0,...,3$;
\item in $D=1+3$ dimensions $m',n'=0,...,3$;
\item in $D=1+2$ dimensions $m,n=0,1,2$;
\end{itemize}
\item for the membrane coordinate indices $\underline i,\underline j=\tau,\sigma,y$;
\item for the string coordinate indices $i,j=\tau,\sigma$;
\item for the spinor indices
\begin{itemize}
\item in $D=10,11$ dimensions $\hat\alpha,\hat\beta,\hat\gamma=1,...,32$;
\item in $D=7,8$ dimensions $A',B',C'=1,...,8$;
\item in $D=6$ dimensions $A,B,C=1,...,4$;
\item of the $SU(3)$ fundamental representation $a,b,c=1,2,3$;
\item in $D=4,5$ dimensions $\alpha,\beta,\gamma=1,...,4$;
\item in $D=3$ dimensions $\mu,\nu,\lambda=1,2$.
\end{itemize}
\end{itemize}

The definition of the $D=2+3$ $\gamma-$matrices $\gamma^{\underline m}_{\alpha\beta}$ is that adopted in
\cite{0811}. The $D=1+3$ $\gamma-$matrices are defined as
\begin{equation}
\Gamma^{m'}{}_\alpha{}^\beta=\gamma^{m'}{}_\alpha{}^\beta,\ \Gamma^5{}_\alpha{}^\beta=(\Gamma^0\Gamma^1\Gamma^2\Gamma^3)_\alpha{}^\beta:\ \Gamma^5{}_\alpha{}^\gamma\Gamma^5{}_\gamma{}^\beta=-\delta_\alpha^\beta
\end{equation}
and satisfy the Clifford algebra relations
\begin{equation}
\Gamma^{m'}{}_\alpha{}^\gamma\Gamma^{n'}{}_\gamma{}^\beta+\Gamma^{n'}{}_\alpha{}^\gamma\Gamma^{m'}{}_\gamma{}^\beta=2\eta^{m'n'}\delta_\alpha^\beta.
\end{equation}
The antisymmetric $D=1+3$ charge conjugation matrices are
\begin{equation}
C'_{\alpha\beta}=\gamma^{0'}_{\alpha\beta},\ C'^{\alpha\beta}=\tilde\gamma^{0'\alpha\beta}.
\end{equation}
Such definition implies the relation
\begin{equation}
C_{\alpha\beta}=\Gamma^5{}_\alpha{}^\gamma C'_{\gamma\beta},
\end{equation}
where $C_{\alpha\beta}$ is the $Sp(4)$ antisymmetric metric.
It follows that $\Gamma^{m'}_{\alpha\beta}=-\Gamma^{m'}{}_\alpha{}^\gamma C'_{\gamma\beta}$ and the $so(1,3)$ generators
\begin{equation}
\Gamma^{m'n'}_{\alpha\beta}=-\frac12(\Gamma^{m'}{}_\alpha{}^\gamma\Gamma^{n'}{}_\gamma{}^\delta-\Gamma^{n'}{}_\alpha{}^\gamma\Gamma^{m'}{}_\gamma{}^\delta)C'_{\delta\beta}
\end{equation}
are symmetric in the spinor indices, while other antisymmetrized
products of $4d$ $\gamma-$matrices are antisymmetric.

In terms of $D=1+2$ real symmetric $\gamma-$matrices
$\sigma^m_{\mu\nu}$, $\tilde\sigma^{m\mu\nu}$ and antisymmetric
metric tensors $\varepsilon_{\mu\nu}$, $\varepsilon^{\mu\nu}$
introduced in \cite{0811} the $4d$ $\gamma-$matrices and $so(1,3)$ generators are
expressed as
\begin{equation}\label{4dto3d}
\Gamma^{m}{}_\alpha{}^\beta=-\left(\begin{array}{cc}
\sigma^m{}_\mu{}^\nu &0\\
0& \sigma^{m\mu}{}_\nu
\end{array}\right),\
\Gamma^{3}{}_\alpha{}^\beta=\left(\begin{array}{cc}
0 &\varepsilon_{\mu\nu}\\
\varepsilon^{\mu\nu}& 0
\end{array}\right),\
\Gamma^{5}{}_\alpha{}^\beta=\left(\begin{array}{cc}
0 &-\varepsilon_{\mu\nu}\\
\varepsilon^{\mu\nu}& 0
\end{array}\right)
\end{equation}
and
\begin{equation}\label{4dto3d'}
\Gamma^{mn}{}_\alpha{}^\beta=\left(\begin{array}{cc}
-\sigma^{mn}{}_\mu{}^\nu &0\\
0& \sigma^{mn}{}_\nu{}^\mu
\end{array}\right),\quad
\Gamma^{3m}{}_\alpha{}^\beta=-\left(\begin{array}{cc}
0& \sigma^m_{\mu\nu}\\
\tilde\sigma^{m\mu\nu}& 0
\end{array}\right),
\end{equation}
where $\sigma^{mn}{}_\mu{}^\nu=\frac12(\sigma^m_{\mu\lambda}\tilde\sigma^{n\lambda\nu}-(m\leftrightarrow n))$.

The $D=7$ $\gamma-$matrices satisfy the following Clifford algebra relations
\begin{equation}
\gamma^{I'}{}_{A'}{}^{C'}\gamma^{J'}{}_{C'}{}^{B'}+\gamma^{J'}{}_{A'}{}^{C'}\gamma^{I'}{}_{C'}{}^{B'}=-2\delta^{I'J'}\delta_{A'}^{B'}
\end{equation}
and are antisymmetric w.r.t. symmetric charge conjugation matrix $C_{A'B'}$ and its inverse $C^{A'B'}$
\begin{equation}
\gamma^{I'}_{A'B'}=\gamma^{I'}{}_{A'}{}^{C'}C_{C'B'}:\ \gamma^{I'}_{A'B'}=-\gamma^{I'}_{B'A'}.
\end{equation}
The $so(7)$ generators are defined as
\begin{equation}\label{so7generators}
\gamma^{I'J'}{}_{A'}{}^{B'}=\frac12(\gamma^{I'}{}_{A'}{}^{C'}\gamma^{J'}{}_{C'}{}^{B'}-\gamma^{J'}{}_{A'}{}^{C'}\gamma^{I'}{}_{C'}{}^{B'})
\end{equation}
and are also antisymmetric in the spinor indices.

From the $6d$ perspective the $D=7$ Majorana spinor is composed of
the 4-component $SU(4)$ spinor and its conjugate
\begin{equation}
\psi_{A'}=\left(\begin{array}{c} \bar\psi_A\\ \psi^A
\end{array}\right)
,\ \bar\psi^{A'}= \left(\begin{array}{c} \psi^A\\ \bar\psi_A
\end{array}\right).
\end{equation}
Accordingly $D=7$ charge conjugation and $\gamma-$matrices have the following representation in terms of the $D=6$ chiral $\gamma-$matrices
$\rho^{I}_{AB}$, $\tilde\rho^{IAB}$ antisymmetric in the spinor indices and $4\times4$ unit matrix
\begin{equation}
C_{A'B'}=-\left(\begin{array}{cc} 0&\delta_A^B\\ \delta^A_B &0
\end{array}\right),\
C^{A'B'}=-\left(\begin{array}{cc} 0&\delta^A_B\\ \delta_A^B &0
\end{array}\right)
\end{equation}
and
\begin{equation}
\gamma^{I}_{A'B'}=\left(\begin{array}{cc} \rho^{I}_{AB}&0\\
0&-\tilde\rho^{IAB}
\end{array}\right),\
\gamma^7_{A'B'}=i\left(\begin{array}{cc} 0&-\delta_A^B\\
\delta^A_B &0
\end{array}\right).
\end{equation}
Then the $so(7)$ generators (\ref{so7generators}) acquire the form
\begin{equation}
\gamma^{IJ}{}_{A'}{}^{B'}=\left(\begin{array}{cc}
-\rho^{IJ}{}_A{}^B &0\\
0& \rho^{IJ}{}_B{}^A
\end{array}\right),\
\gamma^{7I}{}_{A'}{}^{B'}=-i\left(\begin{array}{cc}
0& \rho^{I}_{AB} \\
\tilde\rho^{IAB} &0
\end{array}\right).
\end{equation}

The $D=1+10$ $\gamma-$matrices $\mathfrak{g}^{\hat
m}{}_{\hat\alpha}{}^{\hat\beta}$ that are symmetric w.r.t. antisymmetric charge conjugation matrix $\mathfrak{c}_{\hat\alpha\hat\beta}$ and
satisfy the $D=11$ Clifford algebra relations
\begin{equation}
\mathfrak{g}^{\hat
m}{}_{\hat\alpha}{}^{\hat\gamma}\mathfrak{g}^{\hat
n}{}_{\hat\gamma}{}^{\hat\beta}+\mathfrak{g}^{\hat
n}{}_{\hat\alpha}{}^{\hat\gamma}\mathfrak{g}^{\hat
m}{}_{\hat\gamma}{}^{\hat\beta}=2\eta^{\hat m\hat n}\delta_{\hat\alpha}^{\hat\beta}
\end{equation}
have the following realization in
terms of the $4d$ and $7d$ matrices
\begin{equation}
\mathfrak{g}^{m'}{}_{\hat\alpha}{}^{\hat\beta}=\delta_{A'}^{B'}\Gamma^{m'}{}_\alpha{}^\beta,\
\mathfrak{g}^{I}{}_{\hat\alpha}{}^{\hat\beta}=-\Gamma^5{}_\alpha{}^\beta\gamma^{I}{}_{A'}{}^{B'},\
\mathfrak{g}^{11}{}_{\hat\alpha}{}^{\hat\beta}=-\Gamma^5{}_\alpha{}^\beta\gamma^{7}{}_{A'}{}^{B'}.
\end{equation}
The charge conjugation matrix
$\mathfrak{c}_{\hat\alpha\hat\beta}$ is defined as the direct
product of the $4d$ and $7d$ charge conjugation matrices
\begin{equation}
\mathfrak{c}_{\hat\alpha\hat\beta}=C'_{\alpha\beta}C_{A'B'}.
\end{equation}
Then the $so(1,10)$ generators
\begin{equation}
\mathfrak{g}^{\hat m\hat n}{}_{\hat\alpha}{}^{\hat\beta}=\frac12(\mathfrak{g}^{\hat
m}{}_{\hat\alpha}{}^{\hat\gamma}\mathfrak{g}^{\hat
n}{}_{\hat\gamma}{}^{\hat\beta}-\mathfrak{g}^{\hat
n}{}_{\hat\alpha}{}^{\hat\gamma}\mathfrak{g}^{\hat
m}{}_{\hat\gamma}{}^{\hat\beta})
\end{equation}
split into the $so(1,3)$ generators
$\mathfrak{g}^{m'n'}{}_{\hat\alpha}{}^{\hat\beta}=\delta_{A'}^{B'}\Gamma^{m'n'}{}_\alpha{}^\beta$
and $so(7)$ generators
$\mathfrak{g}^{I'J'}{}_{\hat\alpha}{}^{\hat\beta}=-\delta_{\alpha}^{\beta}\gamma^{I'J'}{}_{A'}{}^{B'}$.
Such realization of the $D=1+10$ $\gamma-$matrices allows to bring
the contribution of the $D=1+10$ translation generators $P_{\hat
m}=(M_{0'm'},V^{8I'})$ to the anticommutator (\ref{antic}) of the $osp(4|8)$ odd
generators to the $(1+10)-$covariant form as
\begin{equation}
\{O_{\alpha A'},O_{\beta B'}\}:\ \frac{i}{2}C_{A'B'}\Gamma^{m'}{}_\alpha{}^\gamma C'_{\gamma\beta}M_{0'm'}-\frac{i}{2}\Gamma^5{}_\alpha{}^\gamma C'_{\gamma\beta}\gamma^{I'}_{A'B'}V^{8I'}=\frac{i}{2}\mathfrak{g}^{\hat
m}{}_{\hat\alpha}{}^{\hat\gamma}\mathfrak{c}_{\hat\gamma\hat\beta}P_{\hat m}.
\end{equation}

\section{$osp(4|8)$ superalgebra}

In analogy with the $osp(4|6)$ superalgebra the (anti)commutation relations of $osp(4|8)$ can be arranged into 5 sets
\begin{eqnarray}
[O_{\alpha\beta},O_{\gamma\delta}]&=&i(C_{\alpha\gamma}O_{\beta\delta}+C_{\alpha\delta}O_{\beta\gamma}+C_{\beta\gamma}O_{\alpha\delta}+C_{\beta\delta}O_{\alpha\gamma}),
\label{so23}\\
\left[O_{A'B'},O_{C'D'}\right]&=&\delta_{A'C'}O_{B'D'}-\delta_{A'D'}O_{B'C'}+\delta_{B'D'}O_{A'C'}-\delta_{B'C'}O_{A'D'},
\label{so8spinor}\\
\{O_{\alpha A'},O_{\beta
B'}\}&=&-C_{A'B'}O_{\alpha\beta}+iC_{\alpha\beta}O_{A'B'},
\label{preantic}\\
 \left[O_{A'B'},O_{\alpha C'}\right]&=&C_{A'C'}O_{\alpha
B'}-C_{B'C'}O_{\alpha A'}, \label{so8fermitrans}\\
 \left[O_{\alpha\beta},O_{\gamma
A'}\right]&=&i(C_{\alpha\gamma}O_{\beta
A'}+C_{\beta\gamma}O_{\alpha A'}).\label{so23fermitrans}
\end{eqnarray}
Note that in the realization of the $osp(4|8)$ superalgebra relevant for the
membrane on $AdS_4\times S^7$ fermionic generators $O_{\alpha A'}$
carry the $8d$ chiral spinor index \cite{Plefka98}.

Eq.~(\ref{so23}) defines the $sp(4)\sim so(2,3)$ algebra
commutation relations that can be cast into the form of $ads_4$ or
$conf_3$ algebra relations \cite{0811}. The difference here is in
the normalization of the $AdS_4$ translation generators $M_{0'm'}$
that implies
\begin{equation}\label{sp4toso23}
O_{\alpha\beta}=-\frac{i}{2}\gamma^{\underline m\underline n}_{\alpha\beta}M_{\underline m\underline n}=-\frac{i}{2}(\Gamma^{m'}{}_\alpha{}^\gamma
C'_{\gamma\beta}M_{0'm'}+\Gamma^{m'n'}{}_\alpha{}^\gamma\Gamma^{5}{}_\gamma{}^\delta
C'_{\delta\beta}M_{m'n'}).
\end{equation}
With the definition of the $3d$ conformal generators
\begin{equation}\label{3dconf}
P_m=-M_{3m}+\frac12M_{0'm},\ K_m=M_{3m}+\frac12M_{0'm},\ D=-M_{0'3}
\end{equation}
the commutation relations of the $3d$ conformal algebra remain the same as in Eq.(B14) of \cite{0811}.

The $so(8)$ commutation relations (\ref{so8spinor})
by the transformation
\begin{equation}\label{sostosov}
O_{A'B'}=-\frac14\gamma^{\underline I\underline
J}_{A'B'}V^{\underline I\underline
J}=-\frac12\gamma^{I'}_{A'B'}V^{8I'}+\frac14\gamma^{I'J'}_{A'B'}V^{I'J'}
\end{equation}
are converted into the vector form
\begin{equation}\label{so8v}
[V^{\underline{IJ}},V^{\underline{KL}}]=\delta^{\underline{IL}}V^{\underline{JK}}-\delta^{\underline{IK}}V^{\underline{JL}}+
\delta^{\underline{JK}}V^{\underline{IL}}-\delta^{\underline{JL}}V^{\underline{IK}}
\end{equation}
that generalizes the commutation relations of the $so(6)$
subalgebra of $osp(4|6)$ \cite{0811}.

In terms of the $D=7$ generators the commutator (\ref{so8v}) decomposes as
\begin{equation}\label{so8in7}
\begin{array}{c}
[V^{8I'},V^{8J'}]=-V^{I'J'},\
[V^{I'J'},V^{8K'}]=\delta^{J'K'}V^{8I'}-\delta^{I'K'}V^{8J'},\\
\left[V^{I'J'},V^{K'L'}\right]=\delta^{I'L'}V^{J'K'}-\delta^{I'K'}V^{J'L'}+\delta^{J'K'}V^{I'L'}-\delta^{J'L'}V^{I'K'}
\end{array}
\end{equation}
or further in terms of the $D=6$ generators
\begin{equation}
\begin{array}{c}
[V^{78},V^{7I}]=-V^{8I},\ [V^{78},V^{8I}]=V^{7I},\\
\left[V^{7I},V^{7J}\right]=[V^{8I},V^{8J}]=-V^{IJ},\
[V^{7I},V^{8J}]=-\delta^{IJ}V^{78},\\
\left[V^{IJ},V^{7(8)K}\right]=\delta^{JK}V^{7(8)I}-\delta^{IK}V^{7(8)J},\\
\left[V^{IJ},V^{KL}\right]=\delta^{IL}V^{JK}-\delta^{IK}V^{JL}+\delta^{JK}V^{IL}-\delta^{JL}V^{IK},
\end{array}
\end{equation}
where the last commutator corresponds to the $so(6)$ algebra.

The $so(8)$ generators decompose into the following $SU(3)$ irreducible parts
\begin{equation}\label{newso8basis}
V^{78},\ V^7{}_{4a},\ V^8{}_{4a},\ V_a{}^4,\
V_a{}^b-\frac13\delta_a^bV_c{}^c,\ V_a{}^a
\end{equation}
and c.c., where we adopted the definition of the $su(4)$ generators
\begin{equation}
V_A{}^B=\frac{i}{4}\rho^{IJ}{}_A{}^BV^{IJ}=\left(
\begin{array}{cc}
V_a{}^b & V_a{}^4\\
V_4{}^b & V_4{}^4\\
\end{array}
\right),\quad V_4{}^4=-V_a{}^a
\end{equation}
and $V^{7(8)}{}_{4a}=V^{7(8)I}\rho^I_{4a}$.

The representation of the 7-sphere isometries in the form suitable for exhibiting its Hopf fibration structure requires the transformation of the $so(8)$ generators (\ref{newso8basis}). The novel generators
\begin{equation}\label{newcp3gen}
T_a=\frac12(V^7{}_{4a}-iV^8{}_{4a}),\quad T^a=-\frac12(V^{74a}+iV^{84a})
\end{equation}
are identified with the coset $su(4)/u(3)$. Their form is dictated by the generation of the $su(4)$ algebra commutation relations and commutativity with the $u(1)$ fiber generator
\begin{equation}\label{deft2}
H=2V_a{}^a-V^{78}.
\end{equation}
The generators
\begin{equation}\label{defu3mod}
\tilde V_a{}^b=V_a{}^b-\frac12\delta_a^bV_c{}^c-\frac14\delta_a^bV^{78}
\end{equation}
are identified with the $u(3)$ subgroup of $su(4)$ and $\tilde V_a{}^a=-\frac12V_a{}^a-\frac34V^{78}$ with the $u(1)$ subgroup
of $u(3)$. 15 Generators (\ref{newcp3gen}), (\ref{defu3mod}) indeed provide the realization of the $su(4)$ algebra
\begin{equation}
\begin{array}{c}
[T_a,T^b]=i(\tilde V_a{}^b+\delta_a^b\tilde V_c{}^c),\ [T_a,\tilde
V_b{}^c]=-i\delta_a^cT_b,\ [T^a,\tilde V_b{}^c]=i\delta_b^aT^c,\\
\left[\tilde V_a{}^b,\tilde V_c{}^d\right]=i(\delta^b_c\tilde
V_a{}^d-\delta^d_a\tilde V_c{}^b)
\end{array}
\end{equation}
with other commutation relations vanishing. Remaining 12
generators from the $so(8)/(su(4)\times u(1))$ coset
\begin{equation}\label{so8su4}
\tilde T_a=-\frac12(V^7{}_{4a}+iV^8{}_{4a}),\quad\tilde T^a=\frac12(V^{74a}-iV^{84a}),\quad V_a{}^4,\quad V_4{}^a
\end{equation}
satisfy the following nonzero commutation relations between themselves and with the above introduced $su(4)$ generators
\begin{equation}
\begin{array}{c}
[T_a,\tilde T_b]=-i\varepsilon_{abc}V_4{}^c,\ [T^a,\tilde
T^b]=i\varepsilon^{abc}V_c{}^4,\\
\left[\tilde
T_a,H\right]=2i\tilde T_a,\ [\tilde
T^a,H]=-2i\tilde T^a,\\
\left[\tilde T_a,\tilde T^b\right]=i(\tilde
V_a{}^b-\delta_a^b(\tilde V_c{}^c+\frac{H}{2})),\ [\tilde T_a,\tilde
V_b{}^c]=\frac{i}{2}\delta_b^c\tilde T_a-i\delta^c_a\tilde T_b,\
[\tilde T^a,\tilde V_b{}^c]=-\frac{i}{2}\delta^c_b\tilde
T^a+i\delta^a_b\tilde T^c,\\
\left[V_4{}^a,V_b{}^4\right]=-i(\tilde V_b{}^a+\delta_b^a(\frac{H}{2}-\tilde V_c{}^c)),\ [V_4{}^a,\tilde V_b{}^c]=-\frac{i}{2}\delta^c_bV_4{}^a+i\delta^a_bV_4{}^c,\
[V_a{}^4,\tilde V_b{}^c]=\frac{i}{2}\delta^c_bV_a{}^4-i\delta^c_aV_b{}^4,\\
\left[V_a{}^4,H\right]=-2iV_a{}^4,\ [V_4{}^a,H]=2iV_4{}^a,\\
\left[V_a{}^4,T_b\right]=-i\varepsilon_{abc}\tilde T^c,\
[V_a{}^4,\tilde
T_b]=-i\varepsilon_{abc}T^c,\\
\left[V_4{}^a,T^b\right]=i\varepsilon^{abc}\tilde T_c,\
[V_4{}^a,\tilde T^b]=i\varepsilon^{abc}T_c.
\end{array}
\end{equation}

The anticommutation relations (\ref{preantic}) allow the following $SO(1,3)\times SO(7)$ covariant representation
\begin{equation}\label{antic}
\begin{array}{rl}
\{O_{\alpha A'},O_{\beta
B'}\}=&\frac{i}{2}C_{A'B'}(\Gamma^{m'}{}_\alpha{}^\gamma
C'_{\gamma\beta}M_{0'm'}+\Gamma^{m'n'}{}_\alpha{}^\gamma\Gamma^{5}{}_\gamma{}^\delta
C'_{\delta\beta}M_{m'n'})\\
-&\frac{i}{4}\Gamma^{5}{}_\alpha{}^\gamma
C'_{\gamma\beta}(2\gamma^{I'}_{A'B'}V^{8I'}-\gamma^{I'J'}_{A'B'}V^{I'J'}).
\end{array}
\end{equation}
The generators $O_{\alpha A'}$ split into the pair of $SU(4)$ chiral spinors
\begin{equation}
O_{\alpha A'}=\left(\begin{array}{c}
\bar O_{\alpha A}\\
O^A_{\alpha}
\end{array}\right)
\end{equation}
and the matrix of the $so(8)$ generators has the following $su(4)$ block structure
\begin{equation}\label{so8matrix}
2\gamma^{I'}_{A'B'}V^{8I'}-\gamma^{I'J'}_{A'B'}V^{I'J'}=2\left(
\begin{array}{cc}
V^8{}_{AB}-iV^7{}_{AB}& i\delta_A^BV^{78}+2iV_A{}^B\\
-i\delta^A_BV^{78}-2iV_B{}^A& -V^{8AB}-iV^{7AB}
\end{array}\right).
\end{equation}
Therefore the anticommutator (\ref{antic}) decomposes as follows
\begin{equation}
\begin{array}{c}
\{O^A_\alpha,O^B_\beta\}=\frac{i}{2}\Gamma^{5}{}_\alpha{}^\gamma C'_{\gamma\beta}(V^{8AB}+iV^{7AB}),\ \{\bar O_{\alpha A},\bar O_{\beta B}\}=-\frac{i}{2}\Gamma^{5}{}_\alpha{}^\gamma C'_{\gamma\beta}(V^8{}_{AB}-iV^7{}_{AB}),\\
\{O^A_\alpha,\bar O_{\beta B}\}=-\frac{i}{2}\delta^A_B(\Gamma^{m'}{}_\alpha{}^\gamma C'_{\gamma\beta}M_{0'm'}+\Gamma^{m'n'}{}_\alpha{}^\gamma\Gamma^{5}{}_\gamma{}^\delta C'_{\delta\beta}M_{m'n'})+\frac{i}{2}\Gamma^{5}{}_\alpha{}^\gamma C'_{\gamma\beta}(i\delta^A_BV^{78}+2iV_B{}^A).
\end{array}
\end{equation}
Further decomposing the fermionic generators into the supersymmetry and superconformal ones
\begin{equation}\label{susygen}
O^A_\alpha=\frac{1}{\sqrt{2}}\left(\begin{array}{c}
Q_{\mu}^A\\
S^{\mu A}
\end{array}\right),\quad\bar O_{\alpha A}=\frac{1}{\sqrt{2}}\left(\begin{array}{c}
\bar Q_{\mu A}\\
\bar S^{\mu}_A
\end{array}\right),
\end{equation}
using the realization of $\Gamma-$matrices in terms of $D=1+2$ ones (\ref{4dto3d}), (\ref{4dto3d'}), the definitions of the $3d$ conformal group generators (\ref{3dconf}) and the $so(8)$ generators (\ref{newcp3gen})-(\ref{defu3mod}), (\ref{so8su4}) we arrive at the set of anticommutators of the $D=3$ $N=8$ superconformal algebra
\begin{equation}
\begin{array}{c}
\{Q^A_\mu,\bar Q_{\nu B}\}=2i\delta^A_B\sigma^m_{\mu\nu}P_m,\ \{S^{\mu A},\bar
S^\nu_B\}=2i\delta^A_B\tilde\sigma^{m\mu\nu}K_m,\\[0.2cm]
\{Q^a_\mu,\bar S^\nu_b\}=-i\delta_\mu^\nu\delta^a_bD+i\delta^a_b\sigma^{mn}{}_\mu{}^\nu M_{mn}-2\delta_\mu^\nu(\tilde V_b{}^a-\delta_b^a\tilde V_c{}^c),\\[0.2cm]
\{\bar Q_{\mu a},S^{\nu b}\}=-i\delta_\mu^\nu\delta^b_aD+i\delta^b_a\sigma^{mn}{}_\mu{}^\nu M_{mn}+2\delta_\mu^\nu(\tilde V_a{}^b-\delta_a^b\tilde V_c{}^c),\\[0.2cm]
\{Q^4_\mu,\bar S^\nu_4\}=-i\delta_\mu^\nu D+i\sigma^{mn}{}_\mu{}^\nu M_{mn}+\delta_\mu^\nu H,\ \{\bar Q_{\mu 4},S^{\nu 4}\}=-i\delta_\mu^\nu D+i\sigma^{mn}{}_\mu{}^\nu M_{mn}-\delta_\mu^\nu H,\\[0.2cm]
\{Q^4_\mu,\bar S^\nu_b\}=-2\delta_\mu^\nu V_b{}^4,\ \{\bar Q_{\mu 4},S^{\nu b}\}=2\delta_\mu^\nu V_4{}^b,\\[0.2cm]
\{Q^a_\mu,\bar S^\nu_4\}=-2\delta_\mu^\nu V_4{}^a,\ \{\bar Q_{\mu a},S^{\nu 4}\}=2\delta_\mu^\nu V_a{}^4,\\[0.2cm]
\{Q^a_\mu,S^{\nu b}\}=2\delta^\nu_\mu\varepsilon^{abc}T_c,\ \{\bar Q_{\mu a},\bar S^\nu_b\}=-2\delta^\nu_\mu\varepsilon_{abc}T^c,\\[0.2cm]
\{Q^a_\mu,S^{\nu 4}\}=2\delta_\mu^\nu\tilde T^a,\ \{\bar Q_{\mu a},\bar S^\nu_4\}=-2\delta_\mu^\nu\tilde T_a,\\[0.2cm]
\{Q^4_\mu,S^{\nu b}\}=-2\delta_\mu^\nu\tilde T^b,\ \{\bar Q_{\mu
4},\bar S^\nu_b\}=2\delta_\mu^\nu\tilde T_b.
\end{array}
\end{equation}
Note that the anticommutators of the $D=3$ $N=6$ superconformal algebra are the same as in \cite{0811} modulo renaming the $su(4)$ generators.

The commutation relations (\ref{so8fermitrans}) of the $so(8)$ generators and the
fermionic generators
can be written in the $7d$ form as
\begin{equation}\label{so8fermitransin7}
[V^{8I'},O_{\alpha A'}]=-\frac12\gamma^{I'}{}_{A'}{}^{B'}O_{\alpha B'},\ [V^{I'J'},O_{\alpha A'}]=\frac12\gamma^{I'J'}{}_{A'}{}^{B'}O_{\alpha B'}
\end{equation}
and further transformed into the commutation relations of the
superconformal generators (\ref{susygen}) with the transformed
$so(8)$ generators (\ref{newcp3gen})-(\ref{defu3mod}),
(\ref{so8su4})
\begin{equation}
\begin{array}{c}
[T^a,Q^b_\mu]=i\varepsilon^{abc}\bar Q_{\mu c},\ [T_a,\bar Q_{\mu b}]=-i\varepsilon_{abc}Q^c_{\mu},\\
\left[T^a,S^{\mu b}\right]=i\varepsilon^{abc}\bar S^\mu_c,\ [T_a,\bar S^{\mu}_{b}]=-i\varepsilon_{abc}S^{\mu c},\\
\left[\tilde T_a,Q^b_\mu\right]=i\delta^b_a\bar Q_{\mu 4},\ [\tilde T^a,\bar Q_{\mu b}]=-i\delta_b^aQ^4_{\mu},\\
\left[\tilde T_a,S^{\mu b}\right]=i\delta^b_a\bar S^{\mu}_{4},\ [\tilde T^a,\bar S^{\mu}_{b}]=-i\delta_b^aS^{\mu 4},\\
\left[\tilde T_a,Q^4_\mu\right]=-i\bar Q_{\mu a},\ [\tilde T^a,\bar Q_{\mu 4}]=iQ_{\mu}^{a},\\
 \left[\tilde T_a,S^{\mu 4}\right]=-i\bar S^{\mu}_{a},\ [\tilde T^a,\bar S^{\mu}_{4}]=iS^{\mu a},\\
\left[H,Q^4_\mu\right]=2iQ^4_\mu,\ [H,\bar Q_{\mu 4}]=-2i\bar
Q_{\mu 4},\ [H,S^{\mu 4}]=2iS^{\mu 4},\
[H,\bar S^{\mu}_{4}]=-2i\bar S^{\mu}_{4},\\
\left[\tilde V_a{}^b,Q^c_\mu\right]=\frac{i}{2}\delta_a^bQ^c_\mu-i\delta_a^cQ^b_\mu,\ [\tilde V_a{}^b,\bar Q_{\mu c}]=-\frac{i}{2}\delta_a^b\bar Q_{\mu c}+i\delta_c^b\bar Q_{\mu a},\\
\left[\tilde V_a{}^b,S^{\mu c}\right]=\frac{i}{2}\delta_a^bS^{\mu
c}-i\delta_a^cS^{\mu b},\ [\tilde V_a{}^b,\bar
S^{\mu}_{c}]=-\frac{i}{2}\delta_a^b\bar
S^{\mu}_{c}+i\delta_c^b\bar S^{\mu}_{a},\\
\left[V_a{}^4,Q^b_\mu\right]=-i\delta_a^bQ^4_\mu,\ [V_4{}^a,\bar Q_{\mu b}]=i\delta^a_b\bar Q_{\mu 4},\\
\left[V_a{}^4,S^{\mu b}\right]=-i\delta_a^bS^{\mu 4},\ [V_4{}^a,\bar S^{\mu}_{b}]=i\delta^a_b\bar S^{\mu}_{4},\\
\left[V_4{}^a,Q^4_\mu\right]=-iQ^a_\mu,\ [V_a{}^4,\bar Q_{\mu 4}]=i\bar Q_{\mu a},\\
\left[V_4{}^a,S^{\mu 4}\right]=-iS^{\mu a},\ [V_a{}^4,\bar
S^\mu_4]=i\bar S^\mu_a.
\end{array}
\end{equation}
Note that the $u(1)$ fiber generator $H$ commutes with the unbroken
supersymmetry generators. This can be considered as its
defining property.

The commutation relations (\ref{so23fermitrans}) of the $sp(4)$ generators and the fermionic generators
in the $AdS_4$ basis acquire the form
\begin{equation}\label{so23fermitransin4}
[M_{0'm'},O_{\alpha
A'}]=-\Gamma_{m'\alpha}{}^\gamma\Gamma^5_\gamma{}^\beta O_{\beta
A'},\ [M_{m'n'},O_{\alpha A'}]=-\frac12\Gamma_{m'n'\alpha}{}^\beta
O_{\beta A'}.
\end{equation}
They can be cast into the commutation relations of the $3d$
conformal generators (\ref{3dconf}) with the fermionic generators
(\ref{susygen}) and have the same form as for the $osp(4|6)$
superalgebra modulo the incorporation of the broken
supersymmetries generators
\begin{equation}
\begin{array}{c}
\left[D,Q^A_\mu\right]=Q^A_\mu,\quad [D,\bar Q_{\mu A}]=\bar
Q_{\mu
A},\\[0.2cm]
\left[M^{mn},Q^A_\mu\right]=\frac12\sigma^{mn}{}_\mu{}^\nu
Q^A_\nu,\quad \left[M^{mn},\bar Q_{\mu
A}\right]=\frac12\sigma^{mn}{}_\mu{}^\nu
\bar Q_{\nu a},\\[0.2cm]
\left[K^m,Q^A_\mu\right]=\sigma^m_{\mu\nu}S^{\nu A},\quad [K^m,\bar Q_{\mu A}]=\sigma^m_{\mu\nu}\bar S^{\nu}_A,\\[0.2cm]
\left[D,S^{\mu A}\right]=-S^{\mu A},\quad [D,\bar S^{\mu}_{A}]=-\bar S^{\mu}_{A},\\[0.2cm]
\left[M^{mn},S^{\mu A}\right]=-\frac12S^{\nu
A}\sigma^{mn}{}_\nu{}^\mu,\quad \left[M^{mn},\bar
S^{\mu}_{A}\right]=-\frac12\bar S^\nu_A
\sigma^{mn}{}_\nu{}^\mu,\\[0.2cm]
\left[P^m,S^{\mu A}\right]=-\tilde\sigma^{m\mu\nu}Q^A_\nu,\quad
[P^m,\bar S^{\mu A}]=-\tilde\sigma^{m\mu\nu}\bar Q_{\nu A}.
\end{array}
\end{equation}

\end{document}